\documentclass[prl,aps,twocolumn,floats,epsfig]{revtex4}
\usepackage{graphicx}
\begin{document}
\title{Calibration of the length of a chain of single gold atoms}
\author{C. Untiedt$^1$, A.\,I. Yanson$^1$, R. Grande$^2$, G.
Rubio-Bollinger$^2$, N. Agra\"{\i}t$^2$, S. Vieira$^2$ and J.\,M.
van Ruitenbeek$^1$} \affiliation{$^1$Kamerlingh Onnes Laboratorium, Leiden University, PO Box 9504, NL-2300 RA Leiden, The Netherlands\\
$^2$Laboratorio de Bajas Temperaturas, Departamento de F\'\i sica de la Materia Condensada C-III, \\
Universidad Aut\'onoma de Madrid, E-28049 Madrid, Spain}
\date{\today}

\begin{abstract}
Using a scanning tunnelling microscope or mechanically
controllable break junction it has been shown that it is possible
to control the formation of a wire made of single gold atoms. In
these experiments an interatomic distance between atoms in the
chain of $\sim$ 3.6 \AA \ was reported which is not consistent
with recent theoretical calculations. Here, using precise
calibration procedures for both techniques, we measure the length
of the atomic chains. Based on the distance between the peaks
observed in the chain length histogram we find the mean value of
the interatomic distance before chain rupture to be 2.5  $ \pm $
0.2 \AA \ . This value agrees with the theoretical calculations
for the bond length. The discrepancy with the previous
experimental measurements was due to the presence of He gas, that
was used to promote the thermal contact, and which affects the
value of the work function that is commonly used to calibrate
distances in scanning tunnelling microscopy and mechanically
controllable break junctions at low temperatures.
\end{abstract}

\pacs{PACS numbers: 62.25.+g., 73.40.Jn, 73.63.Rt}

\maketitle

\section{Introduction}
In the last few years, there has been a significant advance in the
understanding of the electronic properties of atomic-sized
contacts. This has been possible thanks to the use of two
techniques: the scanning tunnelling microscope
(STM)\cite{STM1,STM2} and the mechanically controllable break
junction (MCBJ)\cite{MCBJ}. In both cases the distance between two
electrodes is controlled by means of a piezo-electric transducer
which allows for relative displacements of the electrodes down to
a resolution in the range of picometers.

In these experiments the current that traverses
the contact between two electrodes, at a given bias voltage,
is measured as a function of the relative displacement of these
electrodes. As the contact is broken, the current changes smoothly
during elastic elongation stages, decreasing suddenly in plastic
deformations stages\cite{Forces,Untiedt97}. In the last stage
before breaking the contact, just a few atoms determine the
electronic transport and the conductance is given by the Landauer formula
\begin{equation}
G=\frac{2e^2}{h}\sum_{n=1}^N T_n .
\end{equation}
Here, $N$ is the number of available channels for the electrons
traversing the contact, $e$ is the electron charge, $h$ is
Planck's constant, and $T_n$ is the transmission probability of
the $n$th channel. Just before the contact is broken, when there
is just a single atom at the contact, the conductance for
monovalent metals, such as gold, has been shown\cite{One atom} to
be due to a single conductance channel with transmission
probability close to unity and therefore with a conductance close
to the value $2e^2/h$.

\begin{figure}[!b]
\includegraphics[width=80mm]{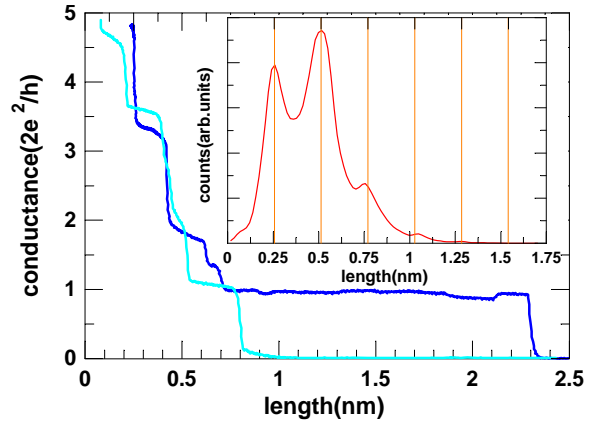}
\caption {\label{plato&histo} Evolution of the conductance for a
gold contact as a function of the interelectrode distance recorded
with the MCBJ technique. The last plateau of the conductance
corresponds to an atomic chain of about six atoms long. The inset
shows a histogram made from 10\,000 recordings of the last plateau
length obtained from individual conductance curves taken at 4.2 K
in vacuum.
  }
\end{figure}

It was observed that occasionally the conductance of the one-atom
contact for gold remains constant while the distance between the
two electrodes increases by more than an interatomic distance, as
it is shown in Fig. \ref{plato&histo}. When it finally  breaks, in
order to make contact again it is necessary to decrease  the
interelectrode distance by the same distance. From such
observations it was concluded that in these cases a wire only one
atom thick was formed between the two  electrodes\cite{chains}.
These wires have also been observed by transmission electron
microscopy (TEM)\cite{TEM,Rodrigues}. The Au-Au bond length was
reported to be 3.6 \AA  \ ($\pm 30\%$) and 3.6--4.0 \AA \ in
\cite{chains} and \cite{TEM} respectively. It was recently shown
that, apart from Au, chain formation can be observed in Pt and Ir
\cite{Smit,Bahn01}. In this paper we concentrate on Au contacts.
Several calculations have confirmed the possibility of the
formation of an atomic chain of gold atoms when stretching the
monatomic
contact\cite{Daneses_sim,Torres_ch,Dani_ch,Landman_sim,daSilva}.
However, in all cases a large discrepancy with the reported
experimental bond length ($\sim$ 3.6 \AA ) was found. The
calculations use different methods including {\it ab initio}
calculations using the local-density approximation
\cite{Torres_ch,Dani_ch,Landman_sim,Okamota,De Maria} and
molecular-dynamics simulations using effective-medium
theory\cite{Daneses_sim} or tight-binding approximations
\cite{daSilva}. For the wire's equilibrium bond length the
different calculations give a distance between 2.32  and 2.55 \AA
, and an upper limit after stretching of 3.0 \AA, much smaller
than the one reported in the experiments.

In this paper we show how the interatomic distance in these atomic
wires can be estimated from the conductance vs electrode
displacement curves. Using this method  we obtain for gold chains
at low temperature (4.2 K) an interatomic distance of 2.5 $\pm$
0.2 \AA \ at the average maximum tensile stress at the moment of
fracture.
\section{Calibration methods}
\label{sec_calib} Since the separation between the peaks in the
length histogram  can provide information on the bond distance in
the chains it is crucial to have a good calibration of the
displacement of the electrodes as a function of the voltage which
is applied to the piezo element of the STM or MCBJ. The various
methods that we have used to calibrate our STM and MCBJ are
described below.
\subsection{Tunnel barrier}
The exponential dependence of the current on the vacuum gap
can be used to make a rough calibration in STM, and until
very recently\cite{Calib_gund}
 it was the only way to obtain a calibration of the interelectrode displacements
in the MCBJ. One makes use of the well-known dependence of the
tunnel current $I_T$ between two electrodes which are separated by
a distance $d$, when a small voltage $V_0$ smaller than the work
function of the electrodes is applied\cite{Expresion de tunel},
\begin{equation}\label{tunnel_curr}
I_T(V_0)= KV_0e^{-2d\sqrt{2m\phi/\hbar^2}},
\end{equation}
where $m$ is the mass of the electron, $\phi$ is the height of the
tunnel barrier, approximately given by the mean value of the work
function of the two electrodes, $K$ is a constant which is related
to the area of the electrodes and to the electronic density of
states at the Fermi level.

\begin{figure}[!t]
\includegraphics[width=80mm]{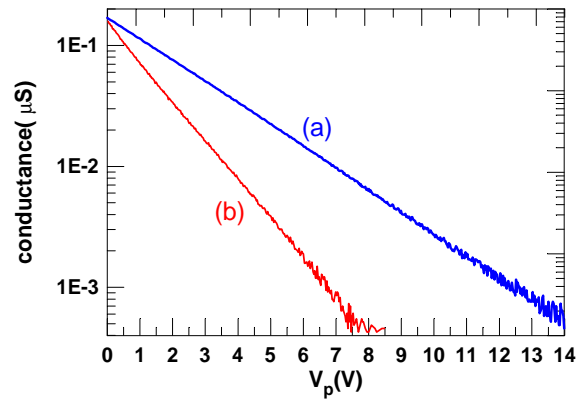}
\caption {\label{tunnel barrier}Exponential dependence of the
current as function of the piezo voltage at a fixed bias voltage
$V$=100 mV when the two electrodes are separated by a tunnel
barrier in vacuum (a) or in a He atmosphere (b).}
\end{figure}

The exponential dependence of the tunnel current with the
interelectrode distance makes it very easy to control that
distance and this is the basis of operation of the STM. If we
represent on a semilog scale the variation of the current as
function of the voltage $V_p$ applied to the piezo element (see
Fig.\ \ref{tunnel barrier}) for the slope $\gamma$ we obtain the
following expression:
\begin{equation}
\gamma = -\frac{\sqrt{2m\phi }}{\hbar } \frac{2\Delta d}{\Delta V_P} .
\end{equation}
This immediately gives us a calibration of the distance as
\begin{equation}\label{cal tunnel}
\kappa=\frac{\Delta d}{\Delta V_p}=\frac{- \hbar \gamma }{2\sqrt{2m\phi }} .
\end{equation}
This expression is very simple and indeed  in experiments with
clean electrodes an exponential behavior of the current as a
function of $V_P$ is found, which would make this a suitable
method for calibration of the response of the system to the
voltage applied to the piezoelectric transducer. Although a more
realistic description for the tunnel barrier must include electron
screening effects, it has been argued\cite{Screening} that these
effects nearly cancel in the logarithmic derivative, at least for
not too small distances.

A problem that arises when using this method is that the value of
the tunnel barrier $\phi$ is dependent on the local work function
of the closest parts of the two electrodes. This local work
function depends mainly on the material with some variation due to
surface distortion and crystal orientation [e.g., the work
function for gold in the (100) direction is 5.47 eV while for the
(111) direction it has a value of 5.31 eV \cite{anisotropia de
wf}]. However, the largest deviation is due to the use of helium.
Helium gas is commonly used to promote thermal contact for cooling
of the STM or MCBJ. It was generally believed that helium gas does
not significantly influence the electron tunneling between two
metallic electrodes. However, very recently it has been found that
atomic layers of adsorbed helium can affect dramatically the work
function measured  with this technique\cite{Influence of He}.
Since the apparent work function was seen to increase for a He
pressure of only 0.01 Torr  by 80\% \ above the clean surface
value, errors of up to 34\%\ are introduced in the distance
calibration due to the presence of a helium atmosphere. For this
reason, in the MCBJ experiments described below we avoid using
helium as a thermal exchange gas.

\subsection{Gundlach oscillations}
A different method for calibrating the MCBJ, based on the Gundlach
oscillations, has been developed by O.Yu.\ Kolesnychenko {\it et
al.} \cite{Calib_gund}. The Gundlach oscillations\cite{Gundlach},
or field-emission resonances, are observed in the tunnel
conductance when a voltage higher than the work function of the
electrodes is applied between them.

\begin{figure}[!t]
\includegraphics[width=80mm]{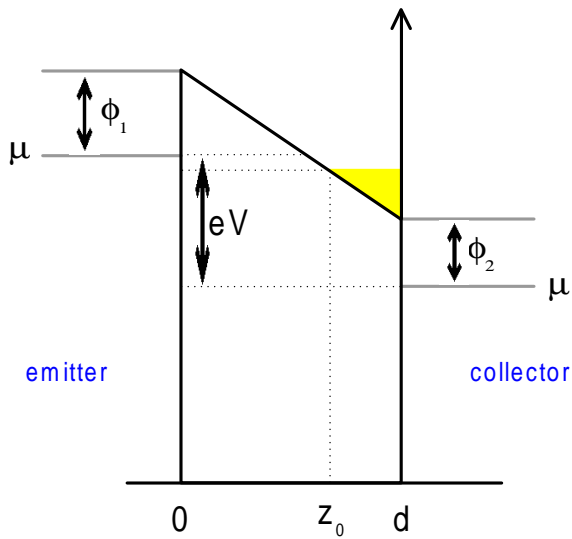}
\caption {\label{diagram}Energy diagram for
field-emission oscillations. Horizontal: $z$ coordinate parallel
to the current direction. Vertical: energy. $\phi_1$ is the work
function for the left electrode and $\phi_2$ that for the right
electrode. The chemical potential for the two electrodes are
shifted by the applied voltage $eV$.}
\end{figure}

As illustrated in Fig. \ref{diagram} when the applied voltage $V$
across the tunnel junction is larger than the work function of the
electrodes, $\phi_{1,2}$, part of the barrier region becomes
classically accessible. In this case the wave function of the
electrons in the region between the electrodes will be determined
by the superposition of the incoming and reflected waves at the
interfaces. This mechanism will give rise to periodic maxima of
the transmission as a function of bias voltage when new resonant
states are formed between the electrodes.
Using the model for this problem proposed by Gundlach\cite{Gundlach},
Kolesnychenko {\it et al.}\
obtained an expression for the differential conductance as a function of
bias voltage given by
\begin{equation}\label{conductance expre}
\frac{dI(V)}{dV}\sim A(V)\cos\left[ \zeta _d(V ) \right].
\end{equation}
 The amplitude of the oscillations, $A(V)$, decreases with voltage as $
V^{-3/2}$ and the
argument for the cosine function is given by
\begin{equation}\label{expr of zeta}
\zeta_d  (V)=\frac{4}{3}\frac{\sqrt{2m}}{\hbar }\frac{(eV-\phi _2)^{3/2}}{eF},
\end{equation}
where $F$ is the electric-field strength in the vacuum gap.

\begin{figure}[!t]
\includegraphics[width=80mm]{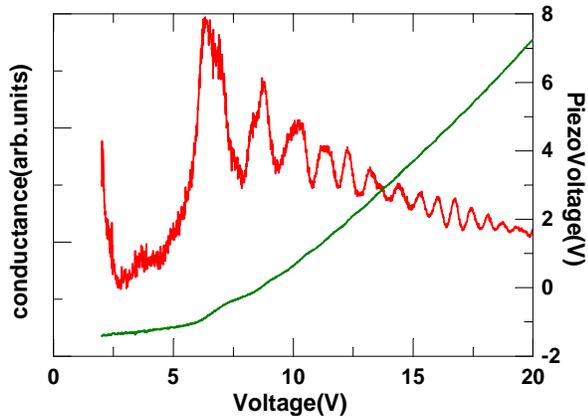}
\caption
{\label{Gundlach_fig}Evolution of conductance and piezo voltage as the applied
voltage across the tunnel junction is increased,
while keeping the current constant.
The analysis of this curve gives a value of 5.3 eV for the tunnel barrier
indicating that the surface is free of adsorbates.}
\end{figure}

The relation between the peak position $V_n$ of the oscillations
in Eq. (\ref{conductance expre}) and their index can be found by
equating Eq.\ (\ref{expr of zeta}) to $2n\pi$:
\begin{equation}\label{expresion gundlach}
eV_n=\phi_2 +\left(\frac{3\pi \hbar }{2\sqrt{2m}} \right)^{2/3}F^{2/3}n^{2/3} .
\end{equation}
During the experiment we keep $F$ constant by applying a feedback to the
piezo voltage in order to
maintain the current constant.
From a plot of $V_n$ versus $n^{2/3}$ the work function $\phi $ is obtained
as the intercept at the voltage axis
and from the slope $\sigma $ of the curve we obtain the field strength
\begin{equation}\label{valor de F}
F=\frac{2\sqrt{2m}}{3\pi \hbar }\sigma ^{3/2} .
\end{equation}
The distance between the two electrodes will then be related to
$F$ and the applied bias voltage according to
\begin{equation}\label{ec_dist}
d=\frac{1}{eF}(eV+\Delta \phi ) ,
\end{equation}
where $\Delta \phi $ is the difference in the work function
between the two electrodes. Using these expressions the procedure
to make the calibration using the Gundlach oscillations will be as
follows: we record the evolution of the conductance, as well as
the piezo voltage $V_p$, as a function of the applied bias voltage
while keeping the current constant (see Fig. \ref{Gundlach_fig}).
Then using Eqs. (\ref{expresion gundlach}) and (\ref{valor de F})
we can calculate the field strength $F\simeq 1.087\; \sigma^{3/2}$
[V/nm]. Finally using Eq. (\ref{ec_dist}) and the response of the
feedback to the voltage changes applied to the junction we can
obtain
\begin{equation}\label{calib gundla}
\kappa =\frac{\Delta d}{\Delta V_p}= \frac{1}{F}\frac{\Delta V}{\Delta V_p}
\end{equation}
for the response at high voltages, where the variation is approximately
linear.
\subsection{Interferometric calibration}
\begin{figure}[!b]
\includegraphics[width=80mm]{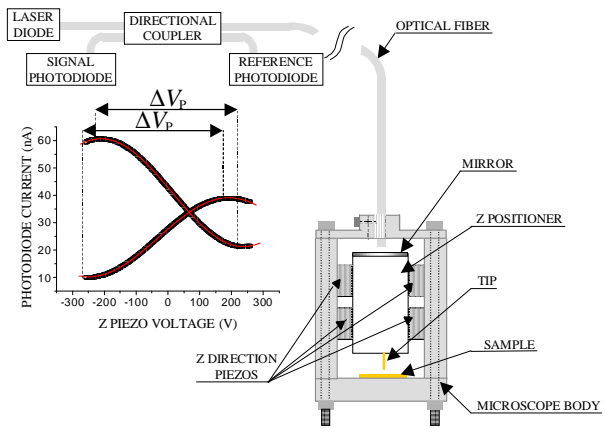}
\caption{\label{STMConfi} Experimental setup for the STM
configuration. The displacement of the tip against the sample is
calibrated by an interferometric method which reliability and
accuracy remain unaffected by the environment. The inset shows two
different traces of the interference pattern measured with
different initial interferometer cavity lengths. }
\end{figure}

The interferometric calibration is a very accurate method for
distance calibration. We have used an all-fiber interferometer
similar to those used in atomic force microscopy \cite{Albrecht} to
calibrate our STM used in the experiments on atomic chains. A
scheme of the experimental set up is shown in Fig. \ref{STMConfi}.
The tip is fastened to a $z$ positioner which is moved by four
stacks of  shear piezos. To calibrate the displacement of the $z$
positioner, the light from a laser diode is focused into a single
mode optical fiber and transmitted through a 2$\times$2
directional coupler which splits the beam. Part of the light is
coupled to a reference photodiode which measures the intensity of
the laser beam. This intensity is the one used as reference when
focusing the light. The remainder of the beam is transmitted to
the end of the fiber which is placed close to a mirror glued to
the rear of the $z$ positioner. In this way an interferometric
cavity is formed between the fiber end and the mirror. About 95\%
of the beam that reaches the fiber end is transmitted, then
reflected at the mirror and directed back into the fiber,
interfering with the beam reflected at the fiber end. The optical
path difference between both beams---twice the interferometric
cavity length---makes the intensity of the resulting beam to be
given by
\begin{equation}\label{ILase}
I = A+B \cos (4 \pi d / \lambda + \delta ),
\end{equation}
where $d$ is the interferometer cavity length, $A$ and $B$ are
constants that decrease with $d$, and $\lambda$ is the wavelength
of the laser beam, 660 nm. The calibration is performed by
approaching the $z$ positioner to the fiber until the intensity
detected by the signal photodiode is sufficiently large. In order
to vary linearly the cavity length, a voltage ramp is applied to
the $z$ direction piezos while the photocurrent of the signal
photodiode is measured. Two typical calibration measurements, with
different initial interferometer cavity lengths (and therefore
with different mean intensities) are shown in the graph in Fig.\
\ref{STMConfi}. For a voltage span $\Delta V_p$, the interference
pattern traces a semiperiod. From Eq. (\ref{ILase}), it follows
that the ratio $\kappa$ between the displacement and the applied
voltage is
\begin{equation}
\kappa =\frac{\Delta d}{\Delta V_p}= \frac{\lambda}{4 \Delta V_p} .
\end{equation}
\section{Experimental results}
\label{sec_result}
For the experiments we have used gold samples of better than
99.99\% \ purity. For the STM experiments we have cleaned the
sample with an H$_2$O-H$_2$SO$_4$ (1:3) solution and mechanically
sharpened the tip, while for the MCBJ a fresh surface was formed
at cryogenic vacuum when breaking the sample. The experiments were
all performed at 4.2 K. The conductance curves, from which the
plateau lengths are obtained, are all measured at a constant bias
voltage of 10 mV.

In Fig.\ \ref{plato&histo} we show a typical experiment were an
atomic chain is formed with the inset showing a histogram of
last-plateau lengths. We have obtained length histograms with both
STM and MCBJ.

A large number of indentation-elongation cycles of gold
nanocontacts was made. Special attention was given to include a
large number of atomic configurations in the statistics, forcing
structural rearrangements of a large number of atoms with frequent
deep indentations of several hundreds of nanometers between
cycles.

In the case of the MCBJ we have measured for several samples
plateau length histograms and each of them was calibrated by both
the tunnel barrier method and by means of the Gundlach
oscillations. For the tunnel barrier method we have taken a work
function for gold of 5.4 eV. In this case the standard deviation
in the distribution of calibration values results into an error of
7\%. Using this calibration we obtain for the interpeak distance
in the length histogram a value of $2.5 \pm 0.2$ \AA. The
calibration using the Gundlach oscillation method was hampered
most of the times by multiple tip effects in the field resonances
and the response of the feedback to the applied voltage often had
a very important quadratic term. Such complication appears to be
characteristic for gold \cite{Influence of He} and the method
works better for most other metals. As a result, we estimate the
error in the calibration to be of the order of 20\% and obtain for
the interpeak distance $2.3 \pm  0.4 $ \AA. We verified that the
calibration obtained by the tunnel barrier method after admitting
He gas into the chamber shifts the peak distance to 3.3 \AA \ in
agreement with \cite{chains}.

\begin{figure}[!t]
\includegraphics[width=80mm]{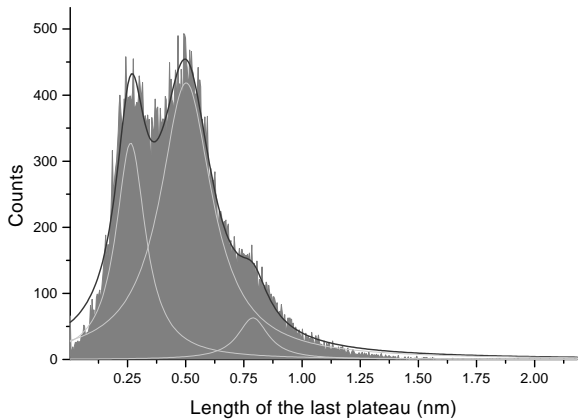}
\caption{\label{HistoSTM} Histogram of lengths for the last
conductance plateau obtained in 65\,000 indentations made with the
STM. We define the length of the last plateau as the distance
between the point where the conductance drops below 1.2 times the
conductance quantum and the one where the contact breaks.  }
\end{figure}

In the case of  the STM configuration, the calibration has been
carried out by the interferometric method. This method has the
advantage of being independent of tip and sample conditions. Using
different lengths of the interferometer cavity, a value of $\kappa
= 3.70\pm 0.13 $ \AA /V is obtained. The experiments were all
performed alternating the conductance with the calibration
measurements every 15\,000 nanocontacts, while the instrument is
maintained at 4.2 K in vacuum. In Fig. \ref{HistoSTM} we show the
resultant length histogram from the STM measurements. We find here
again a preference for contacts with one atom in cross section to
break at specific values of length with a periodicity of $2.6\pm
0.2$ \AA.
\section{Discussion}\label{sec_discuss}
The linear bond between two gold atoms is up to three times
stronger than a bulk bond, as found in experiments and simulations
\cite{Bahn01,Forces_one,Forces on chains}. A single-atom gold
contact can sustain a maximum tensile force of 1.5 nN and before
this limit is reached it is likely that atoms are pulled out of
their position in the banks on either side of the contact. By
repeating such atomic structural changes in the banks the one-atom
contact evolves into a chain several atoms long. The chain finally
breaks when the tensile force necessary to incorporate another
atom from the nearby electrodes into the chain is higher than the
breaking force of the chain itself. There will be characteristic
interelectrode distances for which a chain of $n$ atoms is likely
to break, as we will argue next.

Let us first discuss a length histogram for metals that do not
easily form atomic chains, such as the 4$d$ metals Rh, Pd, and Ag
investigated in Ref.\,\cite{Smit}. For these metals the length
histogram shows only a single peak, usually at a shorter length
than the first peak in the length histogram for Au.  We start
counting the length of the plateau when the conductance drops to a
typical value for a single atom contact, e.g., below 1.2\,$G_0$
for Ag. When pulling further the conductance remains roughly at
this value while the bonds of the atom with the banks and those
inside the banks are being stretched. As soon as the stored
elastic energy reaches a maximum the contact breaks. The breaking
point depends on the local atomic configurations in the banks near
the contact and this leads to a certain width in the peak
distribution. Thermal activation over the breaking barrier  will
also lead to a statistical distribution of observed values. The
peak position in the length histogram shows the most probable
length over which a one-atom contact can be stretched.

For chain-forming metals such as Au the first peak in the length
histogram has the same interpretation as for those that break at a
one-atom contact. Its position is at a longer length reflecting
the stronger bond for low-coordination Au atoms. For all
configurations giving rise to the distribution under the first
peak there are equivalent configurations with the central atom
replaced by two, three, or more atoms, forming a chain. These will
give rise to additional peaks in the length histogram at multiples
of the Au-Au bond distance in the chains, but stretched close to
the breaking point. These distances are the ones at which the
structure reaches the maximum tensile stress while it is not
possible to introduce a new atom into the chain to relax it. If we
consider that the force needed to break an atomic chain, $F_b$, is
independent of the length of the chain\cite{tension_independence}
then the interelectrode distance at which the $n$-atoms chain
breaks can be written as
\begin{equation}
L_n=nL_{at-at}+(n+1)\frac{F_b}{K_a} ,
\end{equation}
were $L_{at-at}$ is the interatomic distance when no tension is
applied and $K_a$ the elastic constant of the bond between atoms
in the chain. Therefore the distance between the peaks in the
plateau length histogram will be constant and equal to $\Delta
=L_{at-at}+ F_b/K_a$, or in other words, equal to the interatomic
distance stretched to the point of breaking.

In this argumentation we have assumed that the banks are not
shortened between the point at which the conductance first is seen
to drop to the one-atom level and the final breaking point. As
long as we limit the discussion to chains of only a few atoms in
length this will be correct since the number of atomic layers in
the banks will not be modified. Note that our value for the bond
distance is based on the first two to four peaks and that atoms
may fold in from both sides. Those events that result in a
significant modification in the structure and effective length of
the banks will only contribute to a smooth background in the
length histogram. Only the chain-forming processes that conserve
the structure of the banks are expected to be responsible for
peaks at regular spacing in the length histogram, and these are
thus expected to correspond to the atom-atom distance in the
chains. The Au-Au distance is measured from the distance between
the peaks in the histogram, and we remark that the position of the
first peak (relative to zero length) can differ from this value.
For Au the first peak is nearly equal to the distance between the
peaks, but different values have indeed been obtained, e.g., in
the case of Pt chain length histograms \cite{Smit}.

The bond distance near the anchoring points of the chain to the
banks are expected to be about 10\% shorter than the bond distance
in the middle of the chain, as illustrated in the calculations by
da Silva {\it et al.} \cite{daSilva}. A small variation in the
bond length is consistent with our data, as can bee seen from the
position of the fourth peak in Fig. \ref{plato&histo}. The fact
that we derive our values for the bond distance mainly from the
first three peaks implies that our result is biased toward the
smaller distances at the anchoring points.

The interatomic distance for gold atomic chains quoted in Ref.
\cite{chains} as 3.6 \AA \ $\pm$ 30\% \ was affected by
calibration inaccuracies in two ways: a systematic error in the
calibration due to He exchange gas condensed onto the gold
surface, resulting in an overestimate of the Au-Au distance. In
addition, variations between various calibrations, which is
reflected in the large uncertainty for the Au-Au distance. We have
now used three independent techniques for an improved calibration
with an accuracy of approximately  7\% for two of them, the
interferometric method for the STM and the tunnel barrier method
for the MCBJ, which show consistence between these techniques and
gives a more accurate value of the bond distance in a chain formed
by gold atoms. The calibration procedure with the largest error
makes use of the Gundlach oscillations and gives a somewhat
smaller, but consistent, value. However, there is still a problem
of consistency with the results of H. Ohnishi {\it et al.}
\cite{TEM} and Rodrigues and Ugarte \cite{Rodrigues} who report an
interatomic distance of about 3.3--4.0 \AA\ obtained from TEM
images. The discrepancy could be due the different methods used to
fabricate the chains or the difference in temperature and
environmental conditions. It has been recently pointed out that
there exists a possibility of incorporating one oxygen atom per
gold atom in an atomic chain \cite{BahnThesis}. This addition
should give rise to a gold-oxygen nanochain with a Au-Au distance
of 3.8 \AA\ for a relaxed chain and up to more than 4.2 \AA\ for a
stretched one. This kind of chain should exhibit a conductance of
one quantum unit.
\section{Conclusions}
We have applied different calibration techniques for the MCBJ and
STM in order to obtain a more accurate value for the distance
between peaks in length histograms of the last plateau of
conductance before rupture of gold contacts at low temperature.
The values obtained for the interatomic distance in a chain of
gold atoms at the point of breaking are $2.5\pm 0.2$, $2.3\pm
0.4$, and $2.6\pm 0.2$ \AA . We obtain an overall value for the
interatomic distance of $2.5 \pm 0.2$ \AA ,which closely agrees
with results from model calculations.
\begin{acknowledgments} \
This work is part of the research program of the ``Stichting
FOM,'' which is financially supported by NWO, it was funded by the
DGI under  contract No. MAT2001-1281 and the research has been
supported by a Marie Curie Fellowship of the European Community
under contract No. HPMF-CT-2000-0072. R.G. acknowledges financial
support from U.A.M. We thank O.I. Shklyarevskii and R.H.M. Smit
for stimulating discussions.
\end{acknowledgments}


\begin{thebibliography}{99}
\bibitem{tension_independence} It has been measured that this is the case
at least for the small chains we consider here and that are no
longer than seven atoms \cite{Forces on chains}.
\end{thebibliography}
\end{document}